# Adversarial Machine Learning and Cybersecurity:

## Risks, Challenges, and Legal Implications

**Authors:**

Micah Musser*
Jonathan Spring*
Christina Liaghati
Daniel Rohrer
Jonathan Elliott
Rumman Chowdhury
Andrew Lohn*
Ram Shankar Siva Kumar*
Cindy Martinez
Heather Frase
Mikel Rodriguez
Stefan Hermanek
James X. Dempsey*
Brenda Leong
Crystal D. Grant
John Bansemer
Mitt Regan

*Workshop Organizer

## Authors

**Micah Musser** is a research analyst with the CyberAI Project at CSET, where **Andrew Lohn** is a senior fellow. **James X. Dempsey** is senior policy advisor for the Program on Geopolitics, Technology, and Governance, Stanford Cyber Policy Center, and lecturer at the UC Berkeley School of Law. **Jonathan Spring** is a cybersecurity specialist at the Cybersecurity and Infrastructure Security Agency (CISA), and was at the time of the July 2022 workshop an analyst at the CERT Division of the Software Engineering Institute at Carnegie Mellon University. **Ram Shankar Siva Kumar** is a data cowboy at Microsoft Security Research and tech policy fellow at the CITRIS Policy Lab and the Goldman School of Public Policy at UC Berkeley.

**Brenda Leong** is a partner at BNH.ai, a boutique law firm focused on the legal issues surrounding AI. **Christina Liaghati** is AI strategy execution and operations manager for the AI and Autonomy Innovation Center at the MITRE Corporation. **Cindy Martinez** is a policy analyst focusing on AI governance and regulation. **Crystal D. Grant** is a data scientist and geneticist who studies the relationship between emerging technologies and civil liberties. **Daniel Rohrer** is vice president of software product security, focused on advancing architecture and research at NVIDIA. **Heather Frase** and **John Bansemer** both work at CSET, where Heather is a senior fellow leading the AI standards and testing line of research, and John is the director of the CyberAI Project. **Jonathan Elliott** is chief of test and evaluation in the Test and Evaluation Division of the Chief Digital and Artificial Intelligence Office. **Mikel Rodriguez** works on securing AI-enabled systems at DeepMind, and was formerly the director of the AI and Autonomy Innovation Center at the MITRE Corporation at the time of the July 2022 workshop. **Mitt Regan** is McDevitt professor of jurisprudence and co-director of the Center on National Security at Georgetown University Law Center. **Rumman Chowdhury** is the founder of Parity Consulting, an ethical AI consulting group, and was formerly the director of Machine Learning, Ethics, Transparency, and Accountability at Twitter at the time of the July 2022 workshop. **Stefan Hermanek** is a product manager with expertise and experience in AI safety and robustness, and AI red-teaming.

Views expressed in this document do not necessarily represent the views of the U.S. government or any institution, organization, or entity with which the authors may be affiliated.

Reference to any specific commercial product, process, or service by trade name, trademark, manufacturer, or otherwise, does not constitute or imply an endorsement, recommendation, or favoring by the U.S. government, including the U.S. Department of Defense, the Cybersecurity and Infrastructure Security Agency, or any other institution, organization, or entity with which the authors may be affiliated.

## Executive Summary

In July 2022, the Center for Security and Emerging Technology (CSET) at Georgetown University and the Program on Geopolitics, Technology, and Governance at the Stanford Cyber Policy Center convened a workshop of experts to examine the relationship between vulnerabilities in artificial intelligence systems and more traditional types of software vulnerabilities. Topics discussed included the extent to which AI vulnerabilities can be handled under standard cybersecurity processes, the barriers currently preventing the accurate sharing of information about AI vulnerabilities, legal issues associated with adversarial attacks on AI systems, and potential areas where government support could improve AI vulnerability management and mitigation.

Attendees at the workshop included industry representatives in both cybersecurity and AI red-teaming roles; academics with experience conducting adversarial machine learning research; legal specialists in cybersecurity regulation, AI liability, and computer-related criminal law; and government representatives with significant AI oversight responsibilities.

This report is meant to accomplish two things. First, it provides a high-level discussion of AI vulnerabilities, including the ways in which they are disanalogous to other types of vulnerabilities, and the current state of affairs regarding information sharing and legal oversight of AI vulnerabilities. Second, it attempts to articulate broad recommendations as endorsed by the majority of participants at the workshop. These recommendations, categorized under four high-level topics, are as follows:

1. **Topic**: Extending Traditional Cybersecurity for AI Vulnerabilities

    1.1. **Recommendation**: Organizations building or deploying AI models should use a risk management framework that addresses security throughout the AI system life cycle.

    1.2. **Recommendation**: Adversarial machine learning researchers, cybersecurity practitioners, and AI organizations should actively experiment with extending existing cybersecurity processes to cover AI vulnerabilities.

    1.3. **Recommendation**: Researchers and practitioners in the field of adversarial machine learning should consult with those addressing AI bias and robustness, as well as other communities with relevant expertise.

2. **Topic**: Improving Information Sharing and Organizational Security Mindsets

    2.1. **Recommendation**: Organizations that deploy AI systems should pursue information sharing arrangements to foster an understanding of the threat.

    2.2. **Recommendation**: AI deployers should emphasize building a culture of security that is embedded in AI development at every stage of the product life cycle.

    2.3. **Recommendation**: Developers and deployers of high-risk AI systems must prioritize transparency.

3. **Topic**: Clarifying the Legal Status of AI Vulnerabilities

    3.1. **Recommendation**: U.S. government agencies with authority over cybersecurity should clarify how AI-based security concerns fit into their regulatory structure.

    3.2. **Recommendation**: There is no need at this time to amend anti-hacking laws to specifically address attacking AI systems.

4. **Topic**: Supporting Effective Research to Improve AI Security

    4.1. **Recommendation**: Adversarial machine learning researchers and cybersecurity practitioners should seek to collaborate more closely than they have in the past.

    4.2. **Recommendation**: Public efforts to promote AI research should more heavily emphasize AI security, including through funding open-source tooling that can promote more secure AI development.

    4.3. **Recommendation**: Government policymakers should move beyond standards-writing toward providing test beds or enabling audits for assessing the security of AI models.

## Table of Contents

## Introduction

Artificial intelligence (AI) technologies, especially machine learning, are rapidly being deployed in a wide range of commercial and governmental contexts. These technologies are vulnerable to an extensive set of manipulations that can trigger errors, infer private data from training datasets, degrade performance, or disclose model parameters.[1] Researchers have demonstrated major vulnerabilities in numerous AI models, including many that have been deployed in public-facing contexts.[2] As Andrew Moore testified before the U.S. Senate Committee on Armed Services in May 2022, defending AI systems from adversarial attacks is "absolutely the place where the battle's being fought at the moment."[3]

However, AI vulnerabilities may not map straightforwardly onto the traditional definition of a patch-to-fix cybersecurity vulnerability (see the section below on "Extending Traditional Cybersecurity for AI Vulnerabilities"). The differences between AI vulnerabilities and more standard patch-to-fix vulnerabilities have generated ambiguity regarding the status of AI vulnerabilities and AI attacks. This in turn poses a series of corporate responsibility and public policy questions: Can AI vulnerabilities be addressed using traditional methods of cyber risk remediation or mitigation? Are the companies developing and using machine learning products equipped to adequately defend them? What legal liability exists for the developers of AI systems, or for the attackers who undermine them? How can policymakers support the creation of a more secure AI ecosystem?

In July 2022, the Center for Security and Emerging Technology (CSET) at Georgetown University and the Program on Geopolitics, Technology, and Governance at the Stanford Cyber Policy Center convened a workshop of experts to address these questions. Attendees included industry representatives in both cybersecurity and AI red-teaming roles; academics with experience conducting adversarial machine learning research; legal specialists in cybersecurity regulation, AI liability, and computer-related criminal law; and government representatives with significant AI oversight responsibilities. This report summarizes the main takeaways of the workshop, with recommendations for researchers, industry professionals, and policymakers.

**Box 1: Explanation of Key Terms**

Many of the terms used throughout this report can have multiple meanings. For the sake of clarifying our discussion, we use the following terms as described here:

**Artificial intelligence (AI)**: a set of technologies that enable computers to learn to perform tasks traditionally performed by humans. This report uses AI interchangeably with machine learning. There are other approaches to AI research beyond machine learning, but this report focuses on vulnerabilities to machine learning-based models. An important subset of current AI approaches is deep learning, and the field of adversarial machine learning focuses largely on attacking and defending deep learning-based models.

**AI system**: a system which includes an AI model as a key component. This definition includes all components of the overall system—including preprocessing software, physical sensors, logical rules, and hardware devices—and can be contrasted with the term "AI model," which we use to refer only to the parameters of the mathematical model produced by an AI training process.

**Vulnerability**: this report adopts the definition of a vulnerability provided by the CERT Guide to Coordinated Vulnerability Disclosure as "a set of conditions or behaviors that allows the violation of an explicit or implicit security policy. Vulnerabilities can be caused by software defects, configuration or design decisions, unexpected interactions between systems, or environmental changes."[4]

**AI vulnerability**: a vulnerability in an AI system, including both vulnerabilities that exploit the mathematical features of AI models, as well as vulnerabilities that arise from the interaction of an AI model with other components of an overall AI system.

**Traditional software vulnerability**: a term used in this report to refer to vulnerabilities in operating systems, workstation applications, server software, and mobile applications with which much of the modern vulnerability management community is most familiar. Example community efforts for traditional software vulnerabilities include CVSSv2 and CVSSv3 and the CVE program.* However, there are many other types of broader cybersecurity vulnerabilities—including vulnerabilities in open-source software, hardware devices, industrial control systems, blockchain protocols, and so on—that may or may not differ from AI vulnerabilities in the ways that this report discusses.

**High-risk AI system**: an AI system that is intended to automate or influence a socially sensitive decision, including those affecting access to housing, credit, employment, healthcare, and so forth. Vulnerabilities—as well as other negative properties, such as bias, unfairness, or discriminatory behavior—are particularly worrisome in these types of systems, as system failures may cause severe harm to individuals. Where considerations affecting the design and deployment of high-risk AI systems are discussed in this report, they should be regarded as minimum requirements for high-risk systems and do not imply endorsement of the use of AI to automate high-risk decision-making in general.

Although the repertoire of attacks studied by adversarial machine learning researchers is expanding, many of these attacks are still focused on lab settings, and a holistic understanding of vulnerabilities in deployed systems is lacking. Due in part to these uncertainties, participants generally shied away from proposing sweeping legal or regulatory changes regarding AI vulnerabilities. At the same time, workshop participants agreed that the risk of attacks on AI systems is likely to grow over time, and that it is important to begin developing mechanisms for addressing AI vulnerabilities now. The fact that computer- and software-based processes already harbor many

*Workshop participants agreed that the risk of attacks on AI systems is likely to grow over time, and that it is important to begin developing mechanisms for addressing AI vulnerabilities now.*

* The Common Vulnerability Scoring System provides a set of properties of a cybersecurity vulnerability that can be used to triage a vulnerability, and a suggested system for ranking vulnerabilities based on those properties. See FIRST, "Common Vulnerability Scoring System SIG," accessed January 30, 2023, https://www.first.org/cvss/. The Common Vulnerabilities and Exposures program aims to provide unique identifiers for all publicly disclosed vulnerabilities. See CVE, https://www.cve.org/, accessed January 29, 2023.

readily exploited vulnerabilities not related to AI should not distract AI developers, AI users, and policymakers from acting now to address AI vulnerabilities.

This report summarizes the general consensus of the workshop related to adversarial machine learning and offers recommendations to improve future responses. These recommendations are divided into four parts:

1. Recommendations regarding the degree to which AI vulnerabilities can be handled under existing cybersecurity processes.

2. Recommendations regarding shifts in organizational culture and information sharing for organizations and individuals actively involved in building AI models, integrating models in business products, and using AI systems.

3. Recommendations regarding the legal issues surrounding AI vulnerabilities.

4. Recommendations regarding future areas of research and government support that can lead to the development of more secure AI systems.

At a high level, we emphasize that—although adversarial machine learning is a complex field with highly technical tools—the problems posed by AI vulnerabilities may be as much social as they are technological. Some of our recommendations emphasize opportunities for industry and government policymakers to promote AI security by investing in technical research. However, the majority of our recommendations focus on changes to processes, institutional culture, and awareness among AI developers and users. While we hope that these recommendations will prompt organizations using AI to proactively think about the security issues surrounding AI systems, we also underscore that individual authors maintain a wide range of viewpoints and do not necessarily endorse each particular recommendation in isolation.

## 1. Extending Traditional Cybersecurity for AI Vulnerabilities

In many senses, attacks on AI systems are not new. Malicious actors have been attempting to evade algorithm-based spam filters or manipulate recommender algorithms for decades. But machine learning models have risen sharply in prevalence over the last decade—including in an increasing number of high-risk contexts.[*] At the same time, researchers have repeatedly demonstrated that vulnerabilities in machine learning algorithms and training processes are pervasive and challenging to remediate.[†] Machine learning-based image and voice recognition systems have been fooled with perturbations imperceptible to humans, datasets have been poisoned in ways that pervert system outputs or render them unreliable, and sensitive data meant to remain private has been reconstructed.[5]

To date, these attacks have mostly occurred in research settings, though there is some evidence of real-world hackers exploiting vulnerabilities in deep learning systems.[6] Moreover, there is a shared expectation that, with the continued incorporation of AI models into a wider range of use cases, the frequency of deep learning-based attacks will grow. Workshop participants suspected that these attacks are likely to be most common wherever there are either clear financial benefits to defeating a machine learning model that would motivate private hackers or strategic advantages to doing so that would motivate a nation-state.

We face a challenge in responding to AI vulnerabilities. On the one hand, existing cybersecurity frameworks are meant to be general enough to cover emerging classes of vulnerabilities such as those generated by deep learning methods. Indeed, it is possible to analyze the risks of AI exploitation under standard risk or vulnerability management frameworks.[7] On the other hand, AI vulnerabilities are distinct from traditional software vulnerabilities in some important respects and may require extensions of, or adjustments to, existing cybersecurity risk governance frameworks. At a high level of abstraction, AI and traditional software vulnerabilities differ in the following ways:

1. AI vulnerabilities generally derive from a complex interaction between training data and the training algorithm. This makes the existence of certain types of vulnerabilities highly dependent on the particular dataset(s) that may be used to train an AI model, often in ways that are difficult to predict or mitigate prior to fully training the model itself. This feature also makes it difficult to test the full space of potential user inputs in order to understand how a system may respond to those inputs.[8]

2. "Patching" a vulnerability in an AI model may require retraining it, potentially at considerable expense, or it may not even be feasible at all.[9] Model retraining to reduce security vulnerabilities may also degrade overall performance on non-malicious system inputs.

3. In many contexts, vulnerabilities in AI systems may be highly ephemeral, as in organizations using continuous training pipelines where models are frequently updated with new data. In other contexts, vulnerabilities may be highly-context dependent, as for instance in organizations that deploy locally fine-tuned versions of a central model across many devices. In either situation, attacks—as well as mitigations—may not transfer well across all versions of the model.

4. There is often deep uncertainty regarding what "counts" as a vulnerability in an AI system. For example, adversarial examples are inputs to an AI system that have been perturbed in some way in order to deliberately degrade the system's performance. But it is hard to distinguish between worrisome "attacks" and neutral—or even expected—user manipulations, like wearing sunglasses to make it harder for facial recognition systems to recognize someone. While this problem is not necessarily unique to AI, it does complicate the matter of defining individual AI vulnerabilities.[10]

---

[*] We note this trend to underscore that all AI systems are likely to carry certain types of AI vulnerabilities that can become particularly worrisome in high-risk contexts. This observation of a trend is not an endorsement of the use of AI in such contexts. Most workshop participants expressed serious reservations about the use of AI in at least some high-risk contexts, but in this report we do not and cannot offer a general rule for evaluating the ethical concerns associated with these use cases.

[†] The field of adversarial machine learning focuses primarily on attacks on deep learning systems. Deep learning is currently the predominant focus of machine learning research, but it is important to emphasize that attacks on deep learning-based models are in some ways elaborations of attacks on other, more traditional machine learning methods. We do not view the category of "AI vulnerabilities" in deep learning models as different in kind from other types of vulnerabilities, including vulnerabilities in older machine learning models; however, the increasing social impact of deep learning models and the level of research attention paid to them warrants some specific focus on these vulnerabilities. In the remainder of this report, we use "deep learning" to distinguish a subset of relatively newer models (and corresponding attacks) from the full range of machine learning models and their vulnerabilities.

These differences are likely to change how vulnerabilities in AI systems are handled. For instance, if fully "patching" a vulnerability is impossible, AI developers and deployers may be more likely to leave systems with known vulnerabilities online.[11]

Responses will likely focus relatively more on risk mitigation and relatively less on risk remediation, in which the underlying vulnerability is fully removed.

***Recommendations***

**1.1. Organizations building or deploying AI models should use a risk management framework that addresses security throughout the AI system life cycle.** Risk management frameworks are a key element of any organization's cybersecurity policy,[12] and we encourage their use for managing AI security. As with other types of risk management frameworks, it is important for organizations to incorporate them throughout the product development pipeline. AI vulnerabilities will, however, present some unique considerations for risk management frameworks. For instance, if vulnerabilities in machine learning models cannot be as easily patched as many traditional software vulnerabilities, then organizations may opt to mitigate vulnerabilities rather than fix or decommission the model, especially when AI models are a part of a complex system where the removal of one component may result in hard-to-predict changes to the overall system.[*]

Important questions in using risk management frameworks include: How can we assure that this model is reliable and robust? In what contexts is it safe to deploy AI models? What compensating controls are available? How should organizations structure the process of deciding between taking a vulnerable system or feature offline entirely or leaving it in place with mitigations? Are important trade-offs at stake, such as trade-offs between applying defensive measures to a model and ensuring that overall performance remains high? What decisions might be in the best interests of the end-users of a machine learning product, as well as in the best interests of the individuals who will ultimately be affected by it, and is there a way to involve those groups throughout development of the system? In the case of high-risk AI systems, organizations should consider what decisions regarding trade-offs most benefit underrepresented groups, but they should also carefully evaluate whether those groups are harmed by the use of an AI system in the first place.

---

[*] Note that the National Institute of Standards and Technology has, as of 2023, released draft 1.0 of its *Artificial Intelligence Risk Management Framework*. See National Institute of Standards and Technology, "NIST AI 100-1: Artificial Intelligence Risk Management Framework (AI RMF 1.0)," January 23, 2023, https://nvlpubs.nist.gov/nistpubs/ai/NIST.AI.100-1.pdf. This document is "intended for voluntary use and to improve the ability to incorporate trustworthiness considerations into the design, development, use, and evaluation of AI products, services, and systems." We view this document as an important step toward effectively augmenting existing risk management practices to account for the risks posed by AI systems.

**1.2. Adversarial machine learning researchers, cybersecurity practitioners, and AI organizations should actively experiment with extending existing cybersecurity processes to cover AI vulnerabilities.** The cybersecurity community has developed many tools for tracking and mitigating vulnerabilities and for guiding incident response. On the vulnerability management side, these include the Common Vulnerabilities and Exposures (CVE) system for enumerating known vulnerabilities, the Common Vulnerability Scoring System (CVSS) for evaluating the potential risk associated with known vulnerabilities, and the Coordinated Vulnerability Disclosure (CVD) process for coordinating between security researchers and software vendors.[13] Although these processes were not designed with AI systems in mind, there was agreement at the workshop that these mechanisms are likely broad enough for managing many types of AI vulnerabilities. However, more collaboration between cybersecurity practitioners, machine learning engineers, and adversarial machine learning researchers is necessary to appropriately apply them to AI vulnerabilities.[14] Assessing AI vulnerabilities requires technical expertise that is distinct from the skill set of cybersecurity practitioners, and organizations should be cautioned against repurposing existing security teams without additional training and resources.

The differences between AI vulnerabilities and traditional software vulnerabilities might make the use of these processes more complicated. At the same time, workshop participants did not feel that the differences are sufficiently strong to justify creating a separate set of processes to handle AI vulnerabilities.[15] We therefore encourage researchers and organizations to incorporate AI vulnerabilities into established risk management practices.

*More collaboration between cybersecurity practitioners, machine learning engineers, and adversarial machine learning researchers is necessary to appropriately apply [existing cybersecurity processes] to AI vulnerabilities.*

**1.3. Researchers and practitioners in the field of adversarial machine learning should consult with those addressing AI bias and robustness, as well as other communities with relevant expertise.** Multiple workshop participants noted that in some important respects, AI vulnerabilities may be more analogous to other topics such as algorithmic bias than they are to traditional software vulnerabilities. AI fairness researchers have extensively studied how poor data, design choices, and risk decisions have led to model failures that cause real-

world harm; the AI security community should seek to better understand these lessons in developing their own frameworks for evaluating risk and assessing the assurance of AI for use in consequential applications. In general, workshop participants believed that it is important to cultivate greater engagement among adversarial machine learning researchers, the AI bias field, cybersecurity practitioners, other relevant expert groups, and affected communities.

## 2. Improving Information Sharing and Organizational Security Mindsets

Several structural features make it difficult to assess precisely how large the threat of attacks on AI systems is. For one, most information about existing AI vulnerabilities has come from theoretical or academic research settings, from cybersecurity companies, or from internal researchers red-teaming their organizations' AI systems.[16] Second, the absence of a systematic and standardized means for tracking AI assets (such as datasets and models) and their corresponding vulnerabilities makes it difficult to know how widespread vulnerable systems are.[17] And third, for some types of attacks on AI systems, attack detection may require meaningful machine learning or data science expertise to implement, or at least a familiarity with the patterns of behavior that may signal an AI-based attack. Since many cybersecurity teams may not have all the relevant expertise to detect such attacks, organizations may lack the capability—and perhaps the incentive—to identify and disclose AI attacks that do occur.[18]

Even if vulnerabilities are identified or malicious attacks are observed, this information is rarely transmitted to others, whether peer organizations, other companies in the supply chain, end users, or government or civil society observers. Although some potential mechanisms for disseminating information exist,[19] a specialized, trusted forum for incident information sharing on a protected basis is lacking. Several workshop participants from industry and government organizations noted that they would benefit from regular exchanges of information, but that networks for information sharing do not currently exist, and that bureaucratic, policy, and cultural barriers currently inhibit such sharing.

These conditions mean that, under current arrangements, the problem will likely remain mostly unnoticed until long after attackers have successfully exploited vulnerabilities. In order to avoid this outcome, we recommend that organizations developing AI models take significant steps to formalize or make use of information sharing arrangements, to monitor for potential attacks on AI systems, and to foster transparency.

***Recommendations***

**2.1. Organizations that deploy AI systems should pursue information sharing arrangements to foster an understanding of the threat.** Currently, there are few trusted mechanisms for organizations that have observed potential attacks on their AI systems to share that information with others, making it difficult for anyone to understand the scope or nature of the problem. Existing attempts to share information about the risks of AI systems—such as the Artificial Intelligence Incident Database—

rely on public reporting and primarily focus on machine learning failures or misuse rather than intentional manipulation.[20] While these efforts are important, they are best situated for compiling publicly known AI failures, rather than for incentivizing organizations to share information about emerging security threats in a more trusted environment. Mechanisms to encourage more open sharing of information could take a wide range of forms, extending from informal but regular meetings of key industry stakeholders to more formalized structures, repositories, or organizations.[21]

**2.2. AI deployers should emphasize building a culture of security that is embedded in AI development at every stage of the product life cycle.** Many machine learning libraries provide functions that, by default, prioritize processing speed and minor improvements in performance over security.[22] Product teams who only consider security concerns after building their models will likely embed insecurities in the development pipeline for their models, which may be difficult or impossible to remove once models are fully trained.[*] As with all software, organizations should make security a priority in every part of the AI pipeline. This entails providing robust support for adversarial machine learning teams, as well as incorporating those teams in every stage of product development to avoid the problem of "outsourcing" security concerns to a separate team.

*Product teams who only consider security concerns after building their models will likely embed insecurities into the development pipeline for their models, which may be difficult or impossible to remove once models are fully trained.*

**2.3. Developers and deployers of high-risk AI systems must prioritize transparency.** AI models should be assumed to come with inherent vulnerabilities—not to mention other types of failure modes inherent to all statistical models—that are difficult if not impossible to patch. The presumed existence of some of these vulnerabilities in high-risk contexts has significant implications for social well-being and privacy. Given this fact, workshop participants felt that the security features of machine learning models also carry transparency implications. A minimum standard of transparency along these lines might hold that consumers and private citizens should generally be informed when they are being made subject to an AI model in a high-risk context. In addition, where the designers of a model make important decisions about relevant trade-offs—such as those that may exist between security, performance, robustness, or fairness—many participants felt that such decisions should be disclosed to protect end-users or private citizens affected by a model's decisions, as well as to help enable recourse when decisions are harmful or discriminatory.[23] While participants disagreed about just how far this principle of transparency should be taken—with many emphasizing that it should *not* extend to simply disclosing the existence of every vulnerability to the public—the more minimal transparency standards discussed in this section were widely supported.

---

[*] As one example, most machine learning libraries provide default image processing functions that are susceptible to adversarial evasion. While alternative preprocessing functions can easily be used, most libraries make use of the less secure methods by default, and models trained on images that have been preprocessed in one way cannot easily transfer to images preprocessed in a different way without substantial retraining. See Lohn, "Downscaling Attack and Defense." Other security-relevant decisions that are difficult to alter after model training include decisions that affect the security and integrity of training data, the use of other types of input filtering, the choice of upstream models for use in fine-tuning, and so on.

## 3. Clarifying the Legal Status of AI Vulnerabilities

There is no comprehensive AI legislation in the United States (and not likely to be one anytime soon).[24] However, many areas of law—including criminal law, consumer protection statutes, privacy law, civil rights law, government procurement requirements, common law rules of contract, negligence and product liability, and even rules of the U.S. Securities and Exchange Commission regarding the disclosure obligations of publicly-owned companies—are relevant to different aspects of AI. Just as AI fits, albeit uneasily, within traditional cybersecurity risk frameworks, so also is it covered under existing law, but in ways and to a degree that courts and regulators have not yet fully clarified. Much of the policy attention on AI to date has focused on concerns with regard to bias and discrimination. The Federal Trade Commission (FTC), U.S. Equal Employment Opportunity Commission, and Consumer Financial Protection Bureau (among others) have all issued their own guidance regarding the use of AI models in contexts that might violate federal civil rights laws, as well as anti-discrimination and consumer protection laws.[25]

At the state level, the New York State Department of Financial Services has warned insurers that the data they use and the algorithms as well as the predictive models they apply may produce forms of discrimination prohibited by state law.[26] And in California, privacy legislation requires the relatively new California Privacy Protection Agency to adopt regulations "governing access and opt-out rights with respect to businesses' use of automated decisionmaking technology, including profiling and requiring businesses' response to access requests to include meaningful information about the logic involved in those decisionmaking processes."[27]

In keeping with our view that AI vulnerabilities should be handled under existing cybersecurity processes as far as possible, we suggest that AI vulnerabilities are likely best handled by extending and adapting cybersecurity law, not by trying to regulate AI security as an independent topic. Unfortunately, cybersecurity law itself is still evolving and many questions are unsettled or contingent. Requirements vary sector-by-sector, with overlapping federal and state rules. Under the resulting patchwork, protected healthcare data, financial information, software and information systems acquired by the federal government, and critical infrastructure, among other categories of systems or data, face certain cybersecurity requirements imposed by statute or executive order.[28] But there is no comprehensive cybersecurity law that imposes clear statutory obligations on the vast majority of companies. While common law doctrines of negligence, product liability, and contract do apply to AI-based products and systems, legal principles in those fields (including questions about the existence of a duty of care, the economic loss rule in negligence cases, and disclaimers of warranty) mean that almost no cases ever yield a clear ruling on liability for security failings. In federal courts, the standing doctrine further makes it difficult to reach the merits of a claim. The near total absence of cybersecurity cases decided on the merits has stunted the development of clear standards with respect to traditional types of vulnerabilities, let alone those associated with AI.

At the same time, the FTC has claimed that its authority to regulate unfair and deceptive business practices extends to cover businesses that fail to secure customer information with "reasonable" cybersecurity measures.[29] The FTC has brought numerous cases against companies that have failed to secure consumer data. While no case to date directly makes this claim, it is easy to imagine that deploying vulnerable AI systems might trigger similar forms of regulatory oversight, especially where those vulnerabilities stem from features common to all machine learning models that companies can reasonably anticipate. In addition, AI companies that make claims about the robustness and performance of their models could be charged with deceptive practices if those models contain foreseeable vulnerabilities that undermine the company's claims. Through its workshops and nonbinding statements, the FTC has made it clear that it is concerned about the impact of AI.

Although the FTC has brought scores of enforcement actions against companies for failure to protect consumer data, uncertainty hangs over the FTC's authority, especially its assertion that failure to provide reasonable security falls under the unfairness prong of its unfair and deceptive practices jurisdiction.[30] Federal regulatory oversight of AI vulnerabilities would, barring legislative action, similarly need to begin from an ambiguous status of authority.

When it comes to deterring attacks on AI systems, an important law is the federal Computer Fraud and Abuse Act.[31] The CFAA makes it illegal to access information from a computer without authorization (or beyond authorized access), as well as to "damage" a computer by knowingly transmitting a "program, information, code, or command."[32] It has been controversial, especially regarding whether the law applies to the activities of good faith cybersecurity researchers probing systems for vulnerabilities. Those legal risks can be, and for many entities have been, mitigated by vulnerability disclosure programs that authorize, or even invite, independent security researchers to probe a system or product. In terms of the CFAA's application to AI, many of the its provisions turn on whether an attacker has first gained "unauthorized access" or exceeded authorized access to a protected computer, a step that may not be required for many adversarial AI attacks. However, one section of the CFAA makes it illegal to cause damage without authorization to a protected computer, where damages is broadly defined to mean any impairment to the integrity or availability of data, a program, a system, or information.

Our recommendations regarding the legal status of AI vulnerabilities begin from the position that this is a rapidly evolving topic within multiple fields of the law. Workshop participants did not feel that it was appropriate at this time to call for comprehensive legislation to address liability for vulnerabilities in AI systems. Our understanding of the law is still too immature to know whether major changes are needed. At the same time, workshop participants did generally support the following recommendations regarding the legal oversight of AI vulnerabilities.

***Recommendations***

**3.1. U.S. government agencies with authority over cybersecurity should clarify how AI-based security concerns fit into their regulatory structure.** The FTC has issued meaningful guidance on how companies using AI can avoid violating the Fair Credit Reporting Act and the Equal Credit Opportunity Act.[33] A wide number of federal agencies, including the National Institute of Standards and Technology, the Cybersecurity and Infrastructure Security Agency, and the FTC, also provide significant cybersecurity guidance to private industry. While NIST's current efforts to develop an AI Risk Management Framework include some discussion of AI security,[34] current guidance on AI security remains vague and does not articulate concrete risks or discuss appropriate countermeasures.[35] Moreover, despite a 2019 executive order requiring federal agencies to document any potential regulatory authority they might have over

AI systems, many agencies declined to respond or have offered extremely surface-level responses.[36] We encourage agencies with regulatory authority over cybersecurity to articulate more concretely how AI vulnerability fits within that regulatory authority. As part of that effort, agencies should formulate concrete guidance on minimum security standards for AI.

**3.2. There is no need at this time to amend anti-hacking laws to specifically address attacking AI systems.** It is unclear whether some types of attacks studied by adversarial machine learning researchers fit within the main federal anti-hacking law, the CFAA, in large part because they do not require "unauthorized access"—a key phrase in the CFAA—in the way that traditional hacking does.[37] However, any attempt to criminalize "AI hacking" would likely raise thorny overbreadth concerns, similar to those that have plagued the CFAA for years.[38] And while many types of "AI hacking" may not be covered by the CFAA, malicious attacks on AI-based systems are likely already illegal under other laws.[39] For now, the best course is to see how the issues play out in the courts. Therefore, we advise against any attempts to adopt new laws aimed at punishing adversarial machine learning.

## 4. Supporting Effective Research to Improve AI Security

Many of the barriers to developing secure AI systems are social and cultural in nature, not technical. The incentives facing academics, industry professionals, and government researchers all encourage—to some extent—a fixation on marginal improvements in summary performance metrics as the primary sign of progress. While adversarial machine learning is a fast-burgeoning field, by some counts it comprises less than 1 percent of all academic AI research—and the research that does exist is heavily focused toward a small subset of attack types, such as adversarial examples, that may not represent plausible real-world attack scenarios.[40] Security is often a secondary consideration for organizations looking to deploy machine learning models. As long as that remains true, technical interventions can have, at best, a limited ability to make AI systems more secure in general.

At the same time, the research community's level of knowledge about adversarial machine learning remains low. While the number of successful attack strategies explored by researchers has exploded, the feasibility of technically defusing these vulnerabilities is uncertain; in particular, it remains unclear how much general-purpose defense against multiple types of attacks is possible. In this report, we have alluded several times to potential trade-offs between security and performance. Even though the existence of trade-offs is clear, it is difficult to assess their extent, establish options for response, engage all relevant stakeholders in risk management, and characterize the extent to which different goals are being traded against one another. We are not aware of an established process for considering these trade-offs.

While this situation calls for more investment in AI security research generally, it represents a place for government policymakers in particular to make a sizable impact. Security is an area where industry may underinvest, which creates an opportunity for publicly-funded research. Workshop participants felt that funding additional research into AI security should be an important priority, and that policymakers can take a few specific actions to most effectively push this area of research forward.

***Recommendations***

**4.1. Adversarial machine learning researchers and cybersecurity practitioners should seek to collaborate more closely than they have in the past.** As mentioned above, adversarial machine learning research comprises only a very small amount of AI research generally, with the research that does exist focusing heavily on the specific attack vector of adversarial examples.[41] Much research into these topics further focuses on threat scenarios that may be unrealistic—such as by assuming that attackers can

manipulate individual pixels in input images—such that research into more likely types of threat models may currently be receiving insufficient attention. We suggest that further collaboration between adversarial machine learning researchers and cybersecurity researchers could help more effectively identify the most realistic threat scenarios facing AI deployers so they can adequately focus their mitigation efforts.

**4.2. Public efforts to promote AI research should more heavily emphasize AI security, including through funding open-source tooling that can promote more secure AI development.** In recent years, the federal government has dedicated significant funding to AI research under the National AI Initiative Act of 2020, the CHIPS and Science Act of 2022, and numerous agency-specific initiatives or projects.[42] Many of these funds will be used for federal research into AI or disbursed as grants to AI researchers via the National Science Foundation, NIST, the U.S. Department of Energy, and a potential National Artificial Intelligence Research Resource. Much of the rhetoric around these initiatives emphasizes public funding for "basic" AI research, as well as curation of government datasets for specific applications.[43] We suggest that AI security should be viewed as a necessary component of basic AI research. Government policymakers should consider how they can best support security-oriented research that effectively complements the research that private industry is already incentivized to pursue, rather than replicating or competing with it.

In particular, supporting the development of open-source tools that can better help AI engineers incorporate security into their products may be a worthwhile goal. As one workshop participant observed, "Adding a machine learning model [to a product] is two lines of code; adding defenses can take hundreds." Existing machine learning libraries are extensively tooled to support the easy design of hundreds of model architectures, but very few contain significant support for common defensive techniques studied by adversarial machine learning researchers. This is an area with a clear gap in technical tooling that government support could help overcome.

**4.3. Government policymakers should move beyond standards-writing toward providing test beds or enabling audits for assessing the security of AI models.** In recent years, government agencies have begun providing high-level guidance that encourages AI developers to carefully consider the potential security impacts of their models. We welcome these goals and recommend further efforts in this direction,

especially among agencies that may potentially have regulatory authority over the use of AI systems. Yet machine learning is too broad of a field for many of these standards to provide direct answers to questions such as: How do I verify the security of this particular model? Or: What specific techniques apply to this use case to reduce the threat of (for example) membership inference attacks?

While standards-writing is important, we suggest that government policymakers should also engage more actively in providing test beds and audits for AI models.[44] One example of such a program is the Face Recognition Vendor Test, an ongoing voluntary test which facial recognition developers can use to identify potential biases in their models.[45] Similarly, one way for government policymakers to promote research in AI security would be to identify a small number of high-risk scenarios and develop audit tools that could be used by vendors to probe their models for security-relevant vulnerabilities. Such programs would also allow policymakers to develop a better understanding of the degree to which existing AI products are vulnerable to known vectors of attack, while also tracking this level of vulnerability over time.

## Acknowledgments

Our conversations at the July 2022 workshop were deeply enriched by several other participants who could not join in the writing process for this final report. For their contributions to our thinking, we would like to thank Aalok Mehta, lead of U.S. public policy at OpenAI; Cristian Canton Ferrer, engineering leader at Meta's Responsible AI team; and Kendra Albert, public interest technology lawyer and clinical instructor at the Cyberlaw Clinic at Harvard Law School. In addition, we would like to thank Danny Hague for his logistic support in running the July 2022 workshop, as well as Shelton Fitch and Jason Ly for their editorial and design support.

## Endnotes

[1] For an overview of various types of attacks on AI systems, see Elham Tabassi, Kevin J. Burns, Michael Hadjimichael, Andres D. Molina-Markham, and Julian Sexton, "Draft NISTIR 8269: A Taxonomy and Terminology of Adversarial Machine Learning" (National Institute of Standards and Technology, October 2019), https://doi.org/10.6028/NIST.IR.8269-draft. Section 2.1.2., "Techniques," describes the broad classes of attacks studied by adversarial machine learning researchers. For a breakdown of adversarial machine learning that focuses more heavily on techniques at various stages of the "kill chain," see "ATLAS," MITRE, accessed September 27, 2022, https://atlas.mitre.org/.

[2] For a list of some of these incidents, see "Case Studies," MITRE, accessed September 27, 2022, https://atlas.mitre.org/studies/. The AI Incident Database also lists several incidents that are tagged with the label "Intent:Deliberate or expected"; see AI Incident Database, accessed January 18, 2023, https://incidentdatabase.ai/apps/discover?classifications=CSET%3AIntent%3ADeliberate%20or%20expected.

[3] Hearing, "To receive testimony on artificial intelligence applications to operations in cyberspace," before the United States Senate Committee on Armed Services, 117th Congress (2022) (statement of Andrew Moore, vice president and general manager of Cloud AI and Industry Solutions, Google Cloud), 1:46:45, https://www.armed-services.senate.gov/hearings/to-receive-testimony-on-artificial-intelligence-applications-to-operations-in-cyberspace.

[4] Allen D. Householder, "1.2. CVD Context and Terminology Notes," The CERT Guide to Coordinated Vulnerability Disclosure, accessed September 22, 2022, https://vuls.cert.org/confluence/display/CVD/1.2.+CVD+Context+and+Terminology+Notes.

[5] Ian Goodfellow, Nicolas Papernot, Sandy Huang, Rocky Duan, Pieter Abbeel, and Jack Clark, "Attacking Machine Learning with Adversarial Examples," OpenAI Blog, February 24, 2017, https://openai.com/blog/adversarial-example-research/; Ian J. Goodfellow, Jonathon Shlens, and Christian Szegedy, "Explaining and Harnessing Adversarial Examples," arXiv [stat.ML], December 20, 2014, doi.org/10.48550/arXiv.1412.6572; Xinyun Chen, Chang Liu, Bo Li, Kimberly Lu, and Dawn Song, "Targeted Backdoor Attacks on Deep Learning Systems Using Data Poisoning," arXiv [cs.CR], December 15, 2017, doi.org/10.48550/arXiv.1712.05526; and Reza Shokri, Marco Stronati, Congzheng Song, and Vitaly Shmatikov, "Membership Inference Attacks against Machine Learning Models," arXiv [cs.CR], October 18, 2016, doi.org/10.48550/arXiv.1610.05820.

[6] See "Case Studies," MITRE; "ATLAS," MITRE; and Andrew Moore, *Testimony on artificial intelligence.* Some of these emerging in-the-wild attacks may not resemble the sophisticated methods studied by AML researchers. After the release of ChatGPT in November 2022, users quickly began coordinating to develop input prompts that could cause the model to ignore its safety guidelines and output toxic or politically charged content. See, for instance, Michael King, "Meet DAN — The 'Jailbreak' Version of ChatGPT and How to Use it — AI Unchained and Unfiltered," Medium, February 5, 2023,

https://medium.com/@neonforge/meet-dan-the-jailbreak-version-of-chatgpt-and-how-to-use-it-ai-unchained-and-unfiltered-f91bfa679024.

[7] See Jonathan Spring, April Galyardt, Allen D. Householder, and Nathan VanHoudnos, "On managing vulnerabilities in AI/ML systems," *New Security Paradigms Workshop 2020* (October 2020), https://doi.org/10.1145/3442167.3442177; and Andrew Grotto and James X. Dempsey, "Vulnerability Disclosure and Management for AI/ML Systems: A Working Paper with Policy Recommendations," *available at SSRN*, November 15, 2021, https://ssrn.com/abstract=3964084.

[8] There is a whole class of product vulnerabilities defined around separating input data from code; see, for example, "CWE-707: Improper Neutralization," MITRE, accessed January 18, 2023, https://cwe.mitre.org/data/definitions/707.html. However, the tactics for addressing these product vulnerabilities do not generally seem to apply to AI vulnerabilities, even if "input sanitization" as a broad idea may be relevant.

[9] "Patching" an AI vulnerability is difficult to define precisely, but for many types of AI vulnerabilities it is often thought of as entailing some form of model retraining which results in an AI system no longer containing the original vulnerability. However, attempts to formalize this idea are difficult, and in some cases formal definitions of an AI "patch" for a specific type of vulnerability have been shown to fail to guarantee the removal of the vulnerability. See Anvith Thudi, Hengrui Jia, Ilia Shumailov, and Nicolas Papernot, "On the Necessity of Auditable Algorithmic Definitions for Machine Unlearning," arXiv [cs.LG], October 22, 2021, arXiv:2110.11891. Even where model retraining **would** eliminate a vulnerability, it may be financially or computationally prohibitive to do so, and there is rarely any guarantee that model retraining will not introduce further vulnerabilities. For many types of vulnerabilities in composed AI systems, "patching" may be more straightforward and may involve more familiar practices, such as certain types of input filtering.

[10] Existing cybersecurity processes may be flexible enough to account for some ambiguity; for instance, the CVE process generally defers to product owners to define vulnerabilities in their own systems. See "Section 7.1: What Is a Vulnerability," in CVE Numbering Authority (CNA) Rules (CVE Board, March 5, 2020), https://cve.mitre.org/cve/cna/CNA_Rules_v3.0.pdf. But in some cases, identifying the "product owner" of an AI system may be difficult, especially if the training model, "foundation model," or fine-tuned instance of a model are all owned by different organizations and the training algorithm is patented by yet another organization. This concern is analogous to issues in cybersecurity that surround supply chain management. See, for example, "Software Bill of Materials," Cybersecurity and Infrastructure Security Agency, accessed January 18, 2023, https://www.cisa.gov/sbom.

[11] It is worth acknowledging that system administrators already often make similar risk calculations in choosing to leave vulnerable systems online because patching will be difficult or time-consuming. See Frank Li, Lisa Rogers, Arunesh Mathur, Nathan Malkin, and Marshini Chetty, "Keepers of the Machines: Examining How System Administrators Manage Software Updates," in *SOUPS'19: Proceedings of the*

*Fifteenth USENIX Conference on Usable Privacy and Security* (August 12, 2019): https://dl.acm.org/doi/10.5555/3361476.3361496.

[12] Notable or relevant examples of risk management frameworks include Joint Task Force Transformation Initiative, "SP 800-39: Managing Information Security Risk: Organization, Mission, and Information System View" (National Institute of Standards and Technology: March 2011), https://csrc.nist.gov/publications/detail/sp/800-39/final; Brett Tucker, "Advancing Risk Management Capability Using the OCTAVE FORTE Process" (Software Engineering Institute, Carnegie Mellon University: November 2020), http://resources.sei.cmu.edu/library/asset-view.cfm?AssetID=644636; ISO Technical Committee 262 on Risk Management, "ISO 31000:2018 Risk management — Guidelines" (International Organization for Standardization: February 2018), https://www.iso.org/standard/65694.html; Board of Governors of the Federal Reserve System and Office of the Comptroller of the Currency, "SR Letter 11-7: Supervisory Guidance on Model Risk Management," April 4, 2011, https://www.federalreserve.gov/supervisionreg/srletters/sr1107.htm; Ross Anderson, *Security Engineering: A Guide to Building Dependable Distributed Systems, 3rd. Edition* (Wiley, 2020); and Patrick Hall, James Curtis, and Parul Pandey, *Machine Learning for High-Risk Applications: Techniques for Responsible AI* (Sebastopol, CA: O'Reilly, 2022).

[13] CVE, https://www.cve.org/, accessed January 29, 2023; FIRST, "Common Vulnerability Scoring System SIG," accessed January 30, 2023, https://www.first.org/cvss/; Allen D. Householder, Garret Wassermann, Art Manion, and Christopher King, "The CERT Guide to Coordinated Vulnerability Disclosure," (Software Engineering Institute, Carnegie Mellon University: August 2017), https://resources.sei.cmu.edu/library/asset-view.cfm?assetid=503330. An example of a relevant guiding document for incident response processes is Paul Cichonski, Thomas Millar, Tim Grance, and Karen Scarfone, "SP 800-61 Rev. 2: Computer Security Incident Handling Guide" (National Institute of Standards and Technology, August 2012), https://csrc.nist.gov/publications/detail/sp/800-61/rev-2/final.

[14] As one example of cross-disciplinary inspiration, note that some organizations have experimented with "bug bounty" programs to highlight flaws in machine learning models. Although modeled off of "bug bounties" that have traditionally been used to identify security vulnerabilities, these AI bug bounties to date have largely focused on issues like fairness and algorithmic bias. See Kyra Yee and Irene Font Peradejordi, "Sharing learnings from the first algorithmic bias bounty challenge," Twitter Engineering, September 7, 2021, https://blog.twitter.com/engineering/en_us/topics/insights/2021/learnings-from-the-first-algorithmic-bias-bounty-challenge; and "AI Audit Challenge," Stanford Institute for Human-Centered Artificial Intelligence, accessed September 27, 2022, https://hai.stanford.edu/policy/ai-audit-challenge.

[15] One significant topic of conversation focused on how the "unpatchability" of AI vulnerabilities affects the feasibility of CVD processes, which typically operate with the goal of keeping discovered vulnerabilities secret until a patch can be developed and deployed. One participant suggested that unpatchability might make more secrecy societally desirable, as disclosing vulnerabilities without available patches increases their risk of exploitation. However, another participant suggested that end-users should be informed about inherent vulnerabilities in systems they use so that they can apply

mitigations in a timely manner. The unpatchability of AI vulnerabilities therefore instead undermines the case for providing the limited period of secrecy that CVD processes generally include, which are primarily societally beneficial insofar as they incentivize the creation of effective patches. Ultimately, while there was general agreement that some aspects of CVD processes may be challenged by the nature of AI vulnerabilities, there was no consensus on whether or how the basic CVD framework should be changed to accommodate those challenges. It is worth emphasizing that this conversation requires careful nuance in distinguishing between patching a vulnerability, remediating it, and mitigating it: remediation can include decommissioning a system as well as patching it, while mitigating involves reducing the impact or incidence of exploitation rather than removing the underlying vulnerability. See for example DoDI 8531.01, "DoD Vulnerability Management," September 15, 2020, https://www.esd.whs.mil/Portals/54/Documents/DD/issuances/dodi/853101p.pdf, section 3.5. The machine learning engineering community, however, does not generally make these distinctions in discussing vulnerability management. Lack of shared vocabulary may drive misunderstandings and in general will slow down deliberation about the best policy options.

[16] The "Case Studies" compiled by MITRE include a variety of attacks on AI systems. As of September 2022, only one of these cases—that of the Tay chatbot—resembles an in-the-wild attack that matches a standard attack type as studied by adversarial machine learning researchers (in this case, a data poisoning attack). See "Tay Poisoning," MITRE, accessed September 29, 2022, https://atlas.mitre.org/studies/AML.CS0009. Similarly, only one deliberate or malicious incident tracked by the AI Incident Database appears to clearly map onto an in-the-wild attack as studied by the field of adversarial machine learning, again as an example of data poisoning. See Patrick Hall, "Incident 88: Jewish Baby Stroller Image Algorithm," Artificial Intelligence Incident Database, accessed September 27, 2022, https://incidentdatabase.ai/cite/88.

[17] To our knowledge, as of September 2022, only one CVE entry in the National Vulnerability Database discussed a vulnerability in a machine learning model—in this case, a vulnerability arising from the possibility of using a model extraction attack to bypass the spam filtering model used by Proofpoint Email Protection. See "CVE-2019-20634 Detail," *National Vulnerability Database*, accessed September 27, 2022, https://nvd.nist.gov/vuln/detail/CVE-2019-20634.

[18] Several workshop participants noted that they felt that "AI vulnerabilities" are often treated as particularly attention-grabbing—potentially more so than many types of traditional software vulnerabilities—in ways that create strong incentives against disclosure. Especially in absence of established norms about vulnerability disclosure and management, organizations may actively prefer to avoid learning about vulnerabilities in their AI systems and may underinvest in attempts to find such vulnerabilities. This issue is not unique to AI, and although norms around other types of vulnerabilities have gradually emerged, bad policies can sometimes create similar perverse incentives for vulnerability management more generally. For some recent commentary on how certain requirements for software security can generate perverse incentives, see Curtis Kang, "How to Comply With the DoD's Newer and Stricter Software Requirements," *Flashpoint*, August 24, 2022, https://flashpoint.io/blog/department-of-

defense-software-requirements/ and Walter Haydock, "Security Release Criteria," *Deploying Securely*, July 22, 2022, https://haydock.substack.com/p/security-release-criteria.

[19] Currently existing mechanisms include national government CSIRT advisories (such as from CISA, NCSC-NL, or JPCERT/CC), trusted community advisories (such as CERT/CC vulnerability notes), and industry PSIRT bulletins (many companies have well-developed product security teams that regularly issue advisories to their constituents; see "FIRST Teams," FIRST, accessed March 16, 2023, https://www.first.org/members/teams/), as well as the option to simply publish information about identified vulnerabilities in scholarly, personal, or organizational venues. For a high-level overview of various types of information-sharing structures, see *The MITRE Corporation*, "Cyber Information-Sharing Models: An Overview," October 2012, https://www.mitre.org/sites/default/files/pdf/cyber_info_sharing.pdf.

[20] As of September 27, 2022, the AI Incident Database tags only four incidents with the label "Intent:Deliberate or expected." See *AI Incident Database*, accessed September 27, 2022, https://incidentdatabase.ai/apps/discover?classifications=CSET%3AIntent%3ADeliberate%20or%20expected&display=details&incident_id=88&page=1. Of these four incidents, one is clearly a case of malicious data poisoning; one refers to manipulation of trading algorithms resulting in a "flash crash" in 2010 which may or may not have been intended by the "attacker"; one relates to blockchain security and not AI; and one refers to a public deepfake intended to warn the public about the dangers of deepfakes, which was not used to manipulate an AI system. The MITRE list of "Case Studies" for AI attacks is slightly more extensive and includes 15 examples of AI attacks. Although ATLAS continues to accept public contributions of case studies, it is likely to represent a severe undercount of exploits actually being deployed against deep learning systems due to concerns around shining a public spotlight on these exploits.

[21] On the variety of information sharing governance structures and their advantages, see Elaine M. Sedenberg and James X. Dempsey, "Cybersecurity Information Sharing Governance Structures: An Ecosystem of Diversity, Trust, and Tradeoffs," *arXiv [cs.CY]*, May 31, 2018, https://doi.org/10.48550/arXiv.1805.12266. Some workshop participants noted that the features of AI vulnerabilities—such as the "unpatcheability" concerns discussed earlier—could disincentive organizations from publicly discussing discovered vulnerabilities. Other participants noted that this same "unpatcheability" may increase the value of public disclosure, as this feature makes it more important for users to understand the risks they may be accepting in absence of a patch for the vulnerability. We emphasize that there are both technical and cultural barriers posed by AI vulnerabilities that have so far hindered extensive information sharing, but also that many workshop participants did not view these barriers as insurmountable and expressed substantial desire for more openness.

[22] See Andrew Lohn, "Downscaling Attack and Defense: Turning What You See Back Into What You Get," *arXiv [cs.CR]*, https://doi.org/10.48550/arXiv.2010.02456 and Andrew Lohn, "Poison in the Well: Securing the Shared Resources of Machine Learning" (Center for Security and Emerging Technology, June 2021), https://doi.org/10.51593/2020CA013.

[23] Many approaches for increasing transparency around AI product development already exist. See Timnit Gebru, Jamie Morgenstern, Briana Vecchione, Jennifer Wortman Vaughan, Hanna Wallach, Hal Daumé III, and Kate Crawford, “Datasheets for Datasets,” *Communications of the ACM* 64, no. 12 (December 2021): 86–92, https://doi.org/10.1145/3458723 and Margaret Mitchell, Simone Wu, Andrew Zaldivar, Parker Barnes, Lucy Vasserman, Ben Hutchinson, Elena Spitzer, Inioluwa Deborah Raji, and Timnit Gebru, “Model Cards for Model Reporting,” in *FAT '19: Proceedings of the Conference on Fairness, Accountability, and Transparency* (January 2019): 220–229, https://doi.org/10.1145/3287560.3287596. For our purposes, we emphasize that sometimes security-relevant decisions involve balancing multiple competing goals; for instance, adversarial retraining to defend against adversarial attacks may degrade model performance on non-adversarial inputs. We suggest that—in addition to the forms of transparency advocated for by these frameworks—organizations should also consider incorporating information about defensive measures used in training a model, their potential impact on other aspects of the model, and ideally the reasoning behind such decisions. However, deciding what exact information to disclose is a difficult process, as some types of disclosure may enable better-targeted attacks of an AI system; in addition, methods that attempt to increase the auditability of an AI model may themselves be susceptible to adversarial attacks. Dylan Slack, Sophie Hilgard, Emily Jia, Sameer Singh, and Himabindu Lakkaraju, “Fooling LIME and SHAP: Adversarial Attacks on Post hoc Explanation Methods,” *arXiv [cs.LG]*, November 6, 2019, https://arxiv.org/abs/1911.02508.

[24] In October 2022, the White House released a Blueprint for an AI Bill of Rights, self-described as “an exercise in envisioning a future where the American public is protected from the potential harms, and can fully enjoy the benefits, of automated systems." It goes on to say that some of the protections it describes “are already required by the U.S. Constitution or implemented under existing U.S. laws,” but it is not specific in describing those existing laws and it stops short of proposing new legislation. “Blueprint for an AI Bill of Rights: Making Automated Systems Work for the American People,” (The White House Office of Science and Technology Policy, October 22), https://www.whitehouse.gov/wp-content/uploads/2022/10/Blueprint-for-an-AI-Bill-of-Rights.pdf, 8. To date, the most comprehensive proposed legislation on AI in the world is the draft AI Act in the EU, which contains several requirements regarding security, but at a high level of generality. For example, Chapter 2 of the current draft includes the following requirement: “High-risk AI systems shall be resilient as regards attempts by unauthorized third parties to alter their use or performance by exploiting the system vulnerabilities. The technical solutions aimed at ensuring the cybersecurity of high-risk AI systems shall be appropriate to the relevant circumstances and the risks. The technical solutions to address AI specific vulnerabilities shall include, where appropriate, measures to prevent and control for attacks trying to manipulate the training dataset (‘data poisoning’), inputs designed to cause the model to make a mistake (‘adversarial examples’), or model flaws.” Draft Regulation (EU) 2021/0106(COD) of the European Commission of 21 April 2021 on a proposal for a regulation of the European Parliament and of the Council laying down harmonised rules on artificial intelligence (Artificial Intelligence Act) and amending certain Union legislative acts, art. 15(4).

[25] See Andrew Smith, “Using Artificial Intelligence and Algorithms,” FTC: BUSINESS BLOG

(April 8, 2020), https://www.ftc.gov/news-events/blogs/business-blog/2020/04/using-artificial-intelligence-algorithms; U.S. Consumer Financial Protection Bureau, Circular 2022-03, Adverse action notification requirements in connection with credit decisions based on complex algorithms (2022) https://www.consumerfinance.gov/compliance/circulars/circular-2022-03-adverse-action-notification-requirements-in-connection-with-credit-decisions-based-on-complex-algorithms/; U.S. Department of Justice, Algorithms, Artificial Intelligence, and Disability Discrimination in Hiring (2022) https://beta.ada.gov/ai-guidance/; EEOC, The Americans with Disabilities Act and the Use of Software, Algorithms, and Artificial Intelligence to Assess Job Applicants and Employees (2022) https://www.eeoc.gov/laws/guidance/americans-disabilities-act-and-use-software-algorithms-and-artificial-intelligence; and Elisa Jillson, "Aiming for truth, fairness, and equity in your company's use of AI," FTC: BUSINESS BLOG (April 19, 2021), https://www.ftc.gov/news-events/blogs/business-blog/2021/04/aiming-truth-fairness-equity-your-companys-use-ai.

[26] Insurance Circular Letter No. 1 (2019), N. Y. State Dep't Fin. Serv., RE: Use of External Consumer Data and Information Sources in Underwriting for Life Insurance (January 18, 2019), https://dfs.ny.gov/industry_guidance/circular_letters/cl2019_01.

[27] CAL. CIV. CODE, § 1798.185(a)(16).

[28] See, for instance, Exec. Order No. 14028, 3 C.F.R. 556 (2022) and Cyber Incident Reporting for Critical Infrastructure Act of 2022, Pub. L. No. 117-103, §§ 101–107, 136 Stat. 1038–1059 (2022).

[29] See Federal Trade Commission, "FTC Report to Congress on Privacy and Security," September 13, 2021, https://www.ftc.gov/system/files/documents/reports/ftc-report-congress-privacy-security/report_to_congress_on_privacy_and_data_security_2021.pdf. The FTC has even found in the concepts of unfair or deceptive a "de facto" breach disclosure requirement. See Federal Trade Commission, "Security Beyond Prevention: The Importance of Effective Breach Disclosures," May 20, 2022, https://www.ftc.gov/policy/advocacy-research/tech-at-ftc/2022/05/security-beyond-prevention-importance-effective-breach-disclosures.

[30] In FTC v. Wyndham Worldwide Corp., 799 F. 3d 236, 257 (3d Cir. 2015), the Third Circuit held in essence that the FTC's authority did extend to cybersecurity practices, and in LabMD, Inc. v. FTC, 894 F.3d 1221 (11th Cir. 2018) the Eleventh Circuit assumed, but did not expressly hold, that unfairness encompassed negligent cybersecurity practices. However, the Supreme Court's June 30, 2022 ruling in "West Virginia v. Environmental Protection Agency, 597 U.S. ___ (2022)" cast doubt on the authority of the FTC to address data security. In this case, the Court indicated that, "in certain extraordinary cases," regulatory agencies could not issue rules on "major questions" affecting "a significant portion of the American economy" without "clear congressional authorization." While the court was focused on the issuance of rules, its logic may also apply to enforcement actions taken in the absence of rules.

[31] The Computer Fraud and Abuse Act of 1986, 18 U.S.C. § 1030.

[32] For an overview of the CFAA's relevance for adversarial machine learning research, see Ram Shankar Siva Kumar and Kendra Albert, "Smashing the ML Stack for Fun and Lawsuits," talk, Black Hat, filmed August 4, 2021, video of talk, 31:53, https://www.youtube.com/watch?v=e3_4ViYRi20. In Van Buren v. United States, 593 U. S. ___ (2021), the Supreme Court clarified that "unauthorized access" does not include circumstances where a company's terms of service merely forbid certain uses of information on a computer system that the individual is otherwise allowed to access. In May 2022, the U.S. Department of Justice issued a policy on the CFAA stating that attorneys for the government "should decline prosecution if available evidence shows the defendant's conduct consisted of, and the defendant intended, good-faith security research." Press Release, Department of Justice, 9-48.000 - Computer Fraud and Abuse Act (May 19, 2022), https://www.justice.gov/opa/press-release/file/1507126/download.

[33] See Smith, "Using Artificial Intelligence," and Jillson, "Aiming for Truth."

[34] NIST, "AI Risk Management Framework," 14–15.

[35] An exception may be National Highway Traffic Safety Administration, "Cybersecurity Best Practices for the Safety of Modern Vehicles" (U.S. Department of Transportation, September 2022), https://www.nhtsa.gov/sites/nhtsa.gov/files/2022-09/cybersecurity-best-practices-safety-modern-vehicles-2022-tag.pdf. The non-binding document describes NHTSA's guidance to the automotive industry for improving vehicle cybersecurity for safety. It mentions machine learning only once, and uses the phrase "artificial intelligence" not at all, but much of what it recommends is applicable to AI-based systems.

[36] Alex Engler, "The AI Bill of Rights Makes Uneven Progress on Algorithmic Protections," Lawfare, October 7, 2022, https://www.lawfareblog.com/ai-bill-rights-makes-uneven-progress-algorithmic-protections.

[37] Kumar and Albert, "Smashing the ML Stack."

[38] See, for instance, Andrew Crocker and Kurt Opsahl, "Supreme Court Overturns Overbroad Interpretation of CFAA, Protecting Security Researchers and Everyday Users," *Electronic Frontier Foundation*, June 3, 2021, https://www.eff.org/deeplinks/2021/06/supreme-court-overturns-overbroad-interpretation-cfaa-protecting-security.

[39] For example, much attention around malicious adversarial attacks has focused around a 2018 result finding that, with a few pieces of tape—and without requiring "unauthorized access" to a self-driving car's vision recognition system—an autonomous vehicle could be tricked into recognizing a stop sign as a "speed limit 45" sign. Kevin Eykholt, Ivan Evtimov, Earlence Fernandes, Bo Li, Amir Rahmati, Chaowei Xiao, Atul Prakash, Tadayoshi Kohno, and Dawn Song, "Robust Physical-World Attacks on Deep Learning Models," in *2018 IEEE/CVF Conference on Computer Vision and Pattern Recognition (CVPR)* (June 2018), https://doi.org/10.1109/CVPR.2018.00175. If such an attack occurred in the real world and caused physical harm to innocent bystanders, it might be difficult to prosecute it under the CFAA—but

likely not very difficult to prosecute as, say, a form of endangerment. Other AI attacks resulting in financial losses instead of physical harm are similarly likely to be criminalized by other laws than the CFAA. For instance, wire fraud charges have a broader jurisdictional sweep than the CFAA, and could likely apply in many cases where AI attacks result in financial losses, even if the CFAA does not. See Computer Crime and Intellectual Property Section, *Prosecuting Computer Crimes* (Department of Justice Office of Legal Education, 2010), https://www.justice.gov/criminal/file/442156/download, p. 26.

[40] Helen Toner and Ashwin Acharya, "Exploring Clusters of Research in Three Areas of AI Safety" (Center for Security and Emerging Technology, February 2022), https://doi.org/10.51593/20210026. Note that AI safety is broader than AI security alone, although some AI security topics relating to confidentiality issues may not have been captured by this analysis.

[41] Toner and Acharya, "Exploring Clusters of Research."

[42] See National Artificial Intelligence Initiative Act of 2020, 15 U.S.C. §§ 9401–9462 and Research and Development, Competition, and Innovation Act of 2022, Pub. L. 117-167, ———.

[43] For instance, see Daniel E. Ho, Jennifer King, Russell C. Wald, and Christopher Wan, "Building a National AI Research Resource: A Blueprint for the National Research Cloud," (Stanford Institute for Human-Centered Artificial Intelligence: October 2021): https://hai.stanford.edu/sites/default/files/2022-01/HAI_NRCR_v17.pdf.

[44] One initiative in this direction is Guaranteeing AI Robustness Against Deception (GARD), sponsored by DARPA, which resulted in a virtual testbed, toolbox, and benchmarking dataset for use in identifying weaknesses to and defenses against adversarial machine learning methods. *GARD Project*, "Holistic Evaluation of Adversarial Defenses," accessed January 30, 2023, https://www.gardproject.org/.

[45] See National Institute of Standards and Technology, "Face Recognition Vendor Test (FRVT) Ongoing," accessed October 7, 2022, https://www.nist.gov/programs-projects/face-recognition-vendor-test-frvt-ongoing. For the importance of evaluating bias in facial recognition systems, see Joy Buolamwini and Timnit Gebru, "Gender Shades: Intersectional Accuracy Disparities in Commercial Gender Classification," Proceedings of Machine Learning Research 81 (2018): https://proceedings.mlr.press/v81/buolamwini18a/buolamwini18a.pdf.